\documentstyle[12pt,epsf]{article}
\begin{document}

\title{ Lagrangians and Hamiltonians for one-dimensional
autonomous systems } \author{ G. Gonz\'alez  \thanks{E-mail:
gabriel${}_{-}$glez@yahoo.com},
\\{\it Departamento de Matem\'aticas y F{\'\i}sica} \\{\it
I.T.E.S.O.}\\ {\it Perif\'erico Sur \# 8585 C.P. 45090} \\
{\it Guadalajara, Jalisco, M\'exico.}}
\date{}
\maketitle

\begin{abstract}
An equation is obtained to find the Lagrangian for a
one-dimensional autonomous system. The continuity of the first
derivative of its constant of motion is assumed. This equation is
solved for a generic nonconservative autonomous system that has
certain quasi-relativistic properties. A new method based on a
Taylor series expansion is used to obtain the associated
Hamiltonian for this system. These results have the usual
expression for a conservative system when the dissipation
parameter goes to zero. An example of this approach is given.

\end{abstract}

\newpage

\section{Introduction}
Lagrangians and Hamiltonians occupy an important position in the
development of physics. Modern physics theories are formulated in
terms of Hamiltonian or Lagrangian structures. Very often the
Lagrangian of a system can be used to find its constants of motion
which can give insight into the stability and periodicity of the
system \cite{Vu}. For an autonomous system the Hamiltonian itself
is a constant of motion of the system.
\\
When a system is conservative, the Lagrangian and the Hamiltonian
can be obtained by subtracting or adding respectively the kinetic
and potential energy of the system \cite{B1}. The ease with which
such constructions are made has contributed to their immense
popularity in the field of physics. However, this construction is
not useful for finding Lagrangians and Hamiltonians for
nonconservative systems. The reason is that there is not yet a
consistent Lagrangian and Hamiltonian formulation for
nonconservative systems. The problem of obtaining the Lagrangian
and Hamiltonian from the equations of motion of a mechanical
system is a particular case of ``The Inverse problem of the
Calculus of Variations" \cite{Sa}. This topic has been studied by
many mathematicians and theoretical physicists since the end of
the last century. The interest of physicists in this problem has
grown recently because of the quantization of nonconservative
systems. A mechanical system can be quantized once its Hamitonian
is known and this
Hamiltonian is usually obtained from a Lagrangian \cite{DHO}. \\
The problem of the existence of a Lagrangian for one dimensional
systems was solved by Darboux \cite{D}, and the relationship
between the constant of motion and the Lagrangian for one
dimensional autonomous systems was given by
Kobusen-Leubner-L\'opez \cite{JK,CL,GL1}. The problem arises when
one tries to obtain the Hamiltonian expressing the velocity in
terms of the canonical variables, which is not possible to do in
general. The main purpose of this paper is to obtain the
Lagrangian and the Hamiltonian for a nonconservative autonomous
system given by $mdv/dt=(-dU/dx + \gamma (x)v^2)(1 - \alpha^2
v^2)$ where $U(x)$ is the potential energy, $v$ is the velocity,
$\gamma(x)$ is an arbitrary function of position and $\alpha^2$ is
any real number. This mechanical system is of interest because it
represents, at first order of approximation, the motion of a
relativistic particle under the action of a dissipative force
which is proportional to the square of the velocity.

\section{Constant of motion, Lagrangian and Hamiltonian}
Newton's equation of motion for one-dimensional autonomous systems
can be written as the following dynamical system
\begin{equation}
\frac{dx}{dt}=v, \qquad \frac{dv}{dt}=F(x,v), \label{eq1}
\end{equation}
where $x$ is the position of the particle, $v$ is the velocity and
$F(x,v)$ is the force divided by the mass of the particle. Let $K
= K(x,v)$ be a constant of motion of (\ref{eq1}), then
\begin{equation}
v\frac{\partial K}{\partial x}+ F(x,v)\frac{\partial K}{\partial
v}=0. \label{eq2}
\end{equation}
Assuming the following condition over the constant of motion
\begin{equation}
\frac{\partial^2 K}{\partial x \partial v}= \frac{\partial^2
K}{\partial v \partial x}, \label{eq3}
\end{equation}
and using the fact that for one-dimensional autonomous systems a
constant of motion is given in terms of the Lagrangian by
\cite{B1}
\begin{equation}
K(x,v)=v\frac{\partial L}{\partial v}-L, \label{eq4}
\end{equation}
then (\ref{eq3}) leads to
\begin{equation}
v\frac{\partial G}{\partial x} + \frac{\partial (F\,G)}{\partial
v}=0, \label{eq5}
\end{equation}
where the Euler-Lagrange equation \cite{B1} has been used and
$G=\partial^2 L/\partial v^2$. Once a nontrivial solution for $G$
has been found, the Lagrangian is obtained trough the integration
\begin{equation}
L(x,v)=\int\,dv\int\,G(x,v)\,dv+f_{1}(x)v-f_{2}(x), \label{eq6}
\end{equation}
where $f_{1}(x)$ and $f_{2}(x)$ are arbitrary functions. The
second term on the right side of (\ref{eq6}) corresponds to a
gauge of the Lagrangian which brings about an equivalent
Lagrangian \cite{B1}, and it is possible to forget it. The
function $f_{2}(x)$ can be determined as it will be showed for the
following mechanical system given by
\begin{equation}
m\frac{dv}{dt}=(-\frac{dU}{dx}+\gamma (x)v^2)(1-\alpha^2 v^2),
\label{eq7}
\end{equation}
where $U(x)$ is the potential energy,$v$ is the velocity, $\gamma
(x)$ is an arbitrary function of position and $\alpha^2$ is any
real number. It is easy to convince oneself that a solution to
(\ref{eq5}) for the mechanical system (\ref{eq7}) is
\begin{equation}
G(x,v)=\frac{m}{(1-\alpha^2 v^2)^2} \exp\left[-\frac{2}{m}
\left(\int\,\gamma(x)\,dx-\alpha^2 U(x)\right)\right], \label{eq8}
\end{equation}
using (\ref{eq8}) one gets the generalized linear momentum
\begin{equation}
p= \frac{m}{2}\left(\frac{v}{1-\alpha^2
v^2}+\frac{\tanh^{-1}(\alpha v)}{
\alpha}\right)\exp\left[-\frac{2}{m}
\left(\int\,\gamma(x)\,dx-\alpha^2 U(x)\right)\right], \label{eq9}
\end{equation}
and the Lagrangian
\begin{equation}
L(x,v)=\frac{mv}{2\alpha}\tanh^{-1}(\alpha
v)\exp\left[-\frac{2}{m} \left(\int\,\gamma(x)\,dx-\alpha^2
U(x)\right)\right]-f_{2}(x). \label{eq10}
\end{equation}
To determine the function $f_{2}$, one uses the Euler-Lagrange
equation \cite{B1}
\begin{equation}
\frac{d}{dt}\left(\frac{\partial L}{\partial
v}\right)=\frac{\partial L}{
\partial x}.
\label{eq11}
\end{equation}
Substituting (\ref{eq10}) into (\ref{eq11}) one gets the following
equation for $f_{2}(x)$
\begin{equation}
\frac{df_{2}}{dx}=\frac{dU}{dx}\exp\left[-\frac{2}{m}
\left(\int\,\gamma(x)\,dx-\alpha^2 U(x)\right)\right],
\label{eq12}
\end{equation}
integrating equation (\ref{eq12}) with respect to $x$ the function
$f_{2}(x)$ is obtained and the constant of motion is given by
\begin{equation}
K(x,v)=\frac{mv^2}{2(1-\alpha^2 v^2)}\exp\left[-\frac{2}{m}
\left(\int\,\gamma(x)\,dx-\alpha^2 U(x)\right)\right]+f_{2}(x).
\label{eq13}
\end{equation}
To obtain the Hamiltonian of the system, we express the constant
of motion in terms of the position and the generalized momentum
\begin{equation}
H(x,p) \equiv K(x,v(x,p)), \label{eq14}
\end{equation}
therefore one has to solve (\ref{eq9}) for the velocity as a
function of the position and the generalized momentum, which at
first sight may seem like a formidable task but it can be done if
we restrict ourselves to the case $|\alpha v|<1$, as it will be
showed in the following example.
\section{Example}
Consider a relativistic particle of mass at rest $m$ under the
action of a constant force $\lambda>0$ and immersed in a medium
that exerts some type of friction which is proportional to the
square of the velocity. The classical equation of motion for this
system is given by
\begin{equation}
m\frac{dv}{dt}=(\lambda-\gamma v^2)(1-v^2/c^2)^{3/2}, \label{eq15}
\end{equation}
where $\gamma$ is a positive real parameter and $c$ represents the
speed of light. Writing (\ref{eq15}) at first order of
approximation in $v^2/c^2$ we have
\begin{equation}
m\frac{dv}{dt}=(\lambda-\gamma v^2)(1-\alpha^2 v^2), \label{eq16}
\end{equation}
where $\alpha^2=\frac{3}{2c^2}$ . The mechanical system
(\ref{eq16}) is a particular case of the mechanical system given
by (\ref{eq7}) where $U(x)=-\lambda x$ and $\gamma(x)=-\gamma$.
Using (\ref{eq8}), (\ref{eq9}) and (\ref{eq10}) we have
\begin{equation}
G(x,v)=\frac{m}{(1-\alpha^2 v^2)^2}
\exp\left[-\frac{2x}{m}\left(\lambda\alpha^2-\gamma\right)\right],
\label{eq17}
\end{equation}
\begin{equation}
p=\frac{m}{2}\left(\frac{v}{1-\alpha^2
v^2}+\frac{\tanh^{-1}(\alpha v)}{\alpha}\right)
\exp\left[-\frac{2x}{m}\left(\lambda\alpha^2-\gamma
\right)\right], \label{eq18}
\end{equation}
\begin{equation}
L(x,v)=\frac{mv\tanh^{-1}(\alpha v)}{2\alpha}
\exp\left[-\frac{2x}{m}\left(\lambda\alpha^2-\gamma
\right)\right]-f_{2}(x), \label{eq19}
\end{equation}
to find $f_{2}(x)$ we have to solve the equation
\begin{equation}
\frac{df_{2}}{dx}=-\lambda\exp\left[-\frac{2x}{m}
\left(\lambda\alpha^2-\gamma\right)\right], \label{eq20}
\end{equation}
which has the following solution
\begin{equation}
f_{2}(x)=\frac{m\lambda}{2(\lambda\alpha^2-\gamma)}\left(
e^{-2x(\lambda\alpha^2-\gamma)/m} - 1\right). \label{eq21}
\end{equation}
Therefore, one obtains the Lagrangian
\begin{equation}
L(x,v)=\frac{mv\tanh^{-1}(\alpha v)}{2\alpha}
e^{-2x(\lambda\alpha^2-\gamma)/m}-\frac{m\lambda}{2(\lambda\alpha^2-\gamma)}\left(
e^{-2x(\lambda\alpha^2-\gamma)/m} - 1\right), \label{eq22}
\end{equation}
and the constant of motion
\begin{equation}
K(x,v)=\frac{mv^2}{2(1-\alpha^2 v^2)}
e^{-2x(\lambda\alpha^2-\gamma)/m}
+\frac{m\lambda}{2(\lambda\alpha^2-\gamma)}\left(
e^{-2x(\lambda\alpha^2-\gamma)/m} - 1\right). \label{eq23}
\end{equation}
To obtain the Hamiltonian of the system, we restrict ourselves to
the case $|\alpha v|<1$, therefore equation (\ref{eq18}) reads
\begin{equation}
pe^{2x(\lambda\alpha^2-\gamma)/m} =\frac{m}{2}\sum_{n=0}^{\infty}
\left(\frac{2n+2}{2n+1}\right) \alpha^{2n}v^{2n+1}, \label{eq24}
\end{equation}
comparing both sides of (\ref{eq24}) we conclude that
\begin{equation}
v^{2n+1}=\frac{2n+1}{(n+1)!}\left(\frac{ 2\lambda
x}{m}\right)^{n}\frac{p}{m}e^{-2\gamma x/m}, \label{eq26}
\end{equation}
therefore, the Hamiltonian is given by
\begin{equation}
H(x,p)= \frac{m}{2}
e^{-2x(\lambda\alpha^2-\gamma)/m}\sum_{n=0}^{\infty}\alpha^{2n}
v^{2n+2}(x,p) +\frac{m\lambda}{2(\lambda\alpha^2-\gamma)}\left(
e^{-2x(\lambda\alpha^2-\gamma)/m} - 1\right), \label{eq27}
\end{equation}
where $v^{2n+2}(x,p)$ is given by
\begin{equation}
v^{2n+2}(x,p)= \left(\frac{p}{m}e^{-2\gamma
x/m}\frac{2n+1}{(n+1)!}\left(\frac{ 2\lambda
x}{m}\right)^{n}\right)^{(2n+2)/(2n+1)}. \label{eq28}
\end{equation}
Notice that the Hamiltonian (\ref{eq27}) has physical meaning only
when $p>0$ and $x>0$. Using Hamilton's equations of motion we have
\begin{equation}
\dot{x}= e^{-2\lambda\alpha^2 x/m}\sum_{n=0}^{\infty}
\frac{\left(2\lambda\alpha^2 x/m\right)^{n}}
{n!}\left[\frac{(2n+1)pe^{-2\gamma x/m}}
{(n+1)!m}\left(\frac{2\lambda x}{m}\right)^{n}\right]^{1/(2n+1)},
\label{eq29}
\end{equation}
\begin{eqnarray*}
\dot{p}&=& e^{-2x(\lambda\alpha^2-\gamma)/m}
\left[\lambda+(\lambda\alpha^2-\gamma)
\sum_{n=0}^{\infty}\alpha^{2n}\left[\frac{(2n+1)pe^{-2\gamma x/m}}
{(n+1)!m}\left(\frac{2\lambda
x}{m}\right)^{n}\right]^{(2n+2)/(2n+1)}\right] \\
       & & \mbox{} -\frac{2p}{m}e^{-2\lambda \alpha^2
x/m}\sum_{n=0}^{\infty}\frac{\alpha^{2n}}{n!}\left[-\gamma \left(
\frac{2\lambda x}{m}\right)^{n}+\lambda n \left(\frac{2\lambda
x}{m}\right)^{n-1}\right]\left[\frac{(2n+1)pe^{-2\gamma x/m}}
{(n+1)!m}\left(\frac{2\lambda x}{m}\right)^{n}\right]^{1/(2n+1)}.
\label{eq30}
\end{eqnarray*}
Hamilton's first equation of motion (\ref{eq29}) give us the
velocity as a function of the position and the generalized
momentum for the case $|\alpha v|<1$. All the expressions derived
have the right limit when $\gamma
\rightarrow 0$ and $\alpha \rightarrow 0$.  \\

\section{Conclusions}
The general form of the Lagrangian, the constant of motion and the
generalized momentum were obtained for the following
nonconservative autonomous system $m\,dv/dt=(-dU/dx + \gamma(x)
v^2)(1 - \alpha^2v^2)$. The Hamiltonian associated to the system
was found for the case $ |\alpha v|<1 $. All the expressions
obtained in this paper converge to the conservative case
when the dissipation parameter goes to zero. \\

~~~~~\\

\vspace{1cm}


\end{document}